\documentclass[prd, nofootinbib, twocolumn,floatfix,preprintnumbers,superscriptaddress, bibliography]{revtex4-1}
\usepackage[colorlinks, linkcolor=red, anchorcolor=blue, citecolor=blue, colorlinks=true, urlcolor=blue]{hyperref}
\usepackage{amssymb}
\usepackage{amsmath}
\usepackage{amsthm}
\usepackage{mathrsfs}
\usepackage{breakurl}
\usepackage{lipsum}
\usepackage{graphicx}
\usepackage{indentfirst}
\usepackage{subfigure}
\usepackage{pifont}
\usepackage{threeparttable}
\usepackage{bbding}
\usepackage{gensymb}
\usepackage{ulem}
\usepackage{bm}
\usepackage{epstopdf}
\usepackage[figuresright]{rotating}
\usepackage{fontawesome}

\begin{document}

\title{Axion limits from the 10-year gamma-ray emission 1ES\,1215+303} 

\author{Hai-Jun Li} 
\email{lihaijun@itp.ac.cn}
\affiliation{Key Laboratory of Theoretical Physics, Institute of Theoretical Physics, Chinese Academy of Sciences, Beijing 100190, China}
\affiliation{Center for Advanced Quantum Studies, Department of Physics, Beijing Normal University, Beijing 100875, China}

\author{Wei Chao} 
\email{chaowei@bnu.edu.cn}
\affiliation{Center for Advanced Quantum Studies, Department of Physics, Beijing Normal University, Beijing 100875, China}

\author{Yu-Feng Zhou}
\email{yfzhou@itp.ac.cn}
\affiliation{Key Laboratory of Theoretical Physics, Institute of Theoretical Physics, Chinese Academy of Sciences, Beijing 100190, China}
\affiliation{School of Physical Sciences, University of Chinese Academy of Sciences, Beijing 100049, China}
\affiliation{School of Fundamental Physics and Mathematical Sciences, Hangzhou Institute for Advanced Study, UCAS, Hangzhou 310024, China}
\affiliation{International Centre for Theoretical Physics Asia-Pacific, Beijing/Hangzhou, China}

\preprint{ITP-23-254, BNU-23-138}

\date{\today}

\begin{abstract}

We present the limits on photon to axionlike particle (ALP) coupling from the 10-year period observations of the TeV BL Lacertae blazar 1ES\,1215+303 (with redshift $z=0.130$).
The contemporaneous gamma-ray spectra are measured by the collaborations Fermi-LAT and VERITAS with five flux phases from 2008 to 2017, including four low states and one flare.
Using these flux phases, we show the spectral energy distributions (SEDs) under the null/ALP hypotheses and set the combined limit on ALP.
The 95\% $\rm C.L.$ combined limit set by 1ES\,1215+303 with the 10-year gamma-ray data is roughly at the photon-ALP coupling constant $g_{a\gamma} \gtrsim 1.5\times 10^{-11} \rm \, GeV^{-1}$ for the ALP mass $5.0\times10^{-10} \, {\rm eV} \lesssim m_a \lesssim 1.0\times10^{-7} \, {\rm eV}$.

%\pacs{ }
%\keywords{ }

\end{abstract}
\maketitle

\section{Introduction}%%%%%%%%%%%%%%%%%%%Introduction

Axions are well motivated candidates for physics beyond the standard model (BSM) of particle physics.
The QCD axion was first postulated to solve the strong CP problem in the SM \cite{Peccei:1977ur, Peccei:1977hh, Weinberg:1977ma, Wilczek:1977pj}, which simultaneously provides a source for cold dark matter (CDM) through the misalignment mechanism \cite{Preskill:1982cy, Abbott:1982af, Dine:1982ah}.
Additionally, the axionlike particle (ALP), arising in a variety of theories \cite{Arvanitaki:2009fg, Svrcek:2006yi}, is also attractive  CDM candidate \cite{Cadamuro:2011fd, Arias:2012az, Chao:2022blc}, but not associated to the solution of the strong CP problem.

The interaction between the ALP and very high energy (VHE) photon in the astrophysical magnetic fields could lead to some detectable effects, such as a reduced TeV opacity of the Universe \cite{Mirizzi:2007hr, Fairbairn:2009zi, Hooper:2007bq, Simet:2007sa}.
The VHE gamma-ray emissions from the extragalactic sources, $\rm e.g.$, the active galactic nuclei (AGN), are mainly affected by the extragalactic background light (EBL) absorption effect through the electron-positron pair production process $\gamma + \gamma \to e^+ + e^-$.
In this case, the photon-ALP interaction provides a natural mechanism to reduce the EBL absorption \cite{Horns:2012fx, Meyer:2013pny}. 
The common mechanism is considering the photon-ALP conversions and back-conversions in the different astrophysical magnetic fields \cite{Belikov:2010ma, Dominguez:2011xy, DeAngelis:2011id, Horns:2012kw, Wouters:2012qd, Marsh:2021ajy, Li:2022mcf, Carenza:2022zmq, Cao:2023kdu}, see also Ref.~\cite{Galanti:2022ijh} for a review.
If we observe a significant photon-ALP oscillation effect, then the Universe would appear to be more transparent than previously thought based on the pure EBL absorption \cite{HESS:2007xak, MAGIC:2008sib}.
In addition, it also provides a natural mechanism to constrain the ALP properties with the ALP mass $m_a$ and the photon-ALP coupling constant $g_{a\gamma}$ \cite{Abramowski:2013oea, TheFermi-LAT:2016zue, Kohri:2017ljt, Ivanov:2018byi, Zhang:2018wpc, Liang:2018mqm, Libanov:2019fzq, Long:2019nrz, Guo:2020kiq, Li:2020pcn, Li:2021gxs, Cheng:2020bhr, Li:2021zms, Dessert:2022yqq, Jacobsen:2022swa, Li:2022jgi, Mastrototaro:2022kpt, Noordhuis:2022ljw, Li:2022pqa, Pant:2022ibi, Gao:2023dvn, LHAASO:2023lkv, Gao:2023und, Chattopadhyay:2023nuq, Caputo:2023cpv, ciaran_o_hare_2020_3932430}.

In this paper, we investigate the photon-ALP oscillation effects from the TeV blazar 1ES\,1215+303 with the 10-year period gamma-ray observations.
The source 1ES\,1215+303 ($\rm R.A.=12^h17^m52.0819^s$, $\rm Dec.=+30^\circ07'00''635$, J2000), also known as other names, $\rm e.g.$, Ton\,605, ON\,325, B2\,1215+30, and S3\,1215+30, is classified as an intermediate-frequency-peaked BL Lac (IBL) or a high-synchrotron-peaked BL Lac (HBL) object, which was first discovered at VHE by the MAGIC telescopes \cite{MAGIC:2012guq}.
The redshift of this source was measured to be $z=0.1305\pm 0.0030$ \cite{Paiano:2017pol, Furniss:2019tkk}, and we take $z=0.130$ in this work.
Recently, its long-term contemporaneous gamma-ray spectra are measured by the collaborations Fermi-LAT and the Very Energetic Radiation Imaging Telescope Array System (VERITAS) with several low state and flare flux phases from 2008 to 2017 \cite{Fermi-LAT:2020pst}.
Using these 10-year gamma-ray flux phases, we consider the photon-ALP oscillation effects on the gamma-ray spectral energy distributions (SEDs) and present the ALP limits in the $\{m_a, \, g_{a\gamma}\}$ plane. 

This paper is organized as follows.
In Sec.~\ref{sec_source}, we describe the long-term gamma-ray data of 1ES\,1215+303 and show the SEDs without the ALP.
In Sec.~\ref{sec_setup}, we introduce the photon-ALP oscillation effects and set the combined limit on ALP with these gamma-ray spectra.
The conclusion is given in Sec.~\ref{sec_conclusion}.

\begin{figure*}[t]%%%%%%%%%%%%%%%%%%SEDs
\centering
\includegraphics[width=0.335\textwidth]{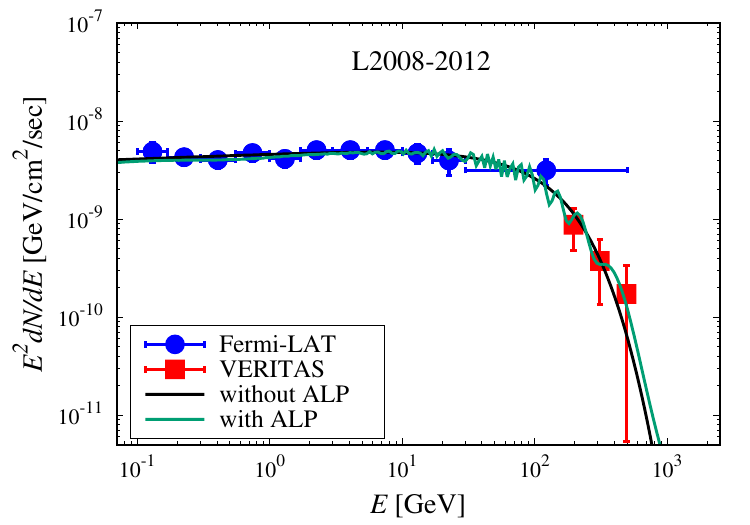}\includegraphics[width=0.335\textwidth]{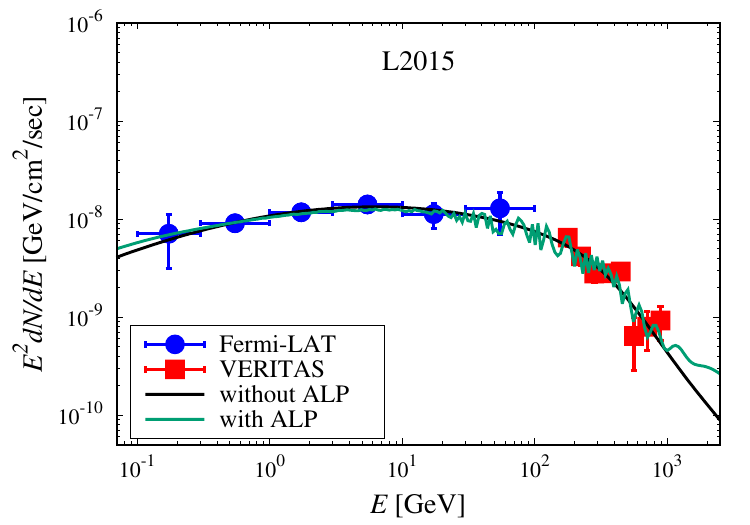}
\includegraphics[width=0.335\textwidth]{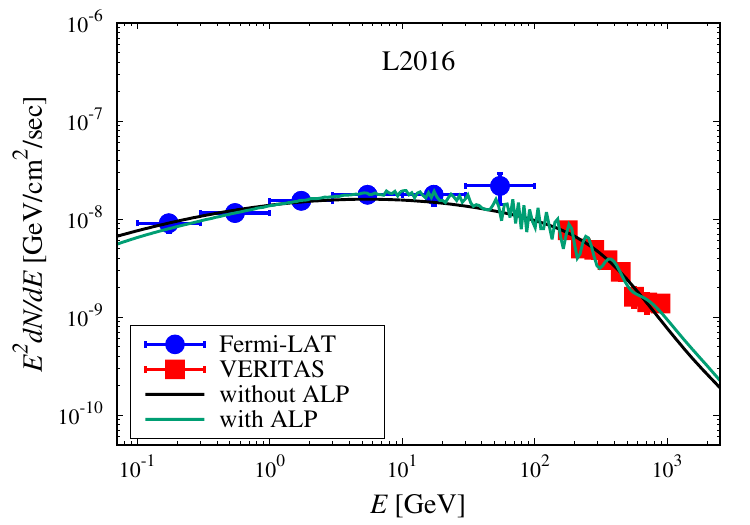}\includegraphics[width=0.335\textwidth]{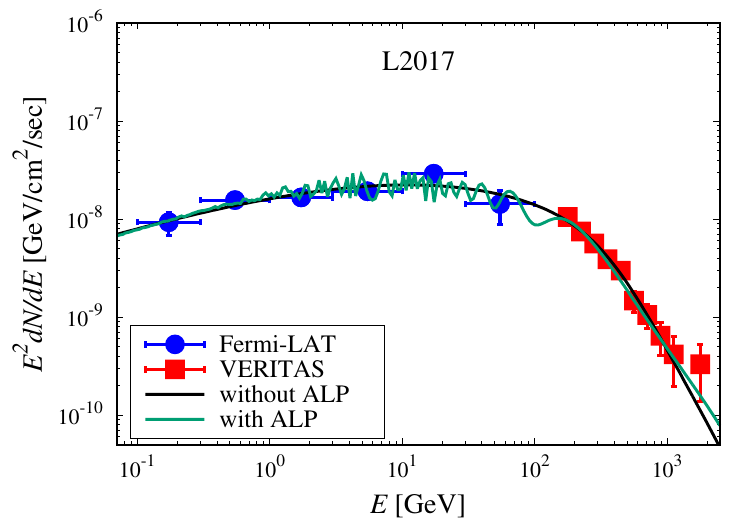}\includegraphics[width=0.335\textwidth]{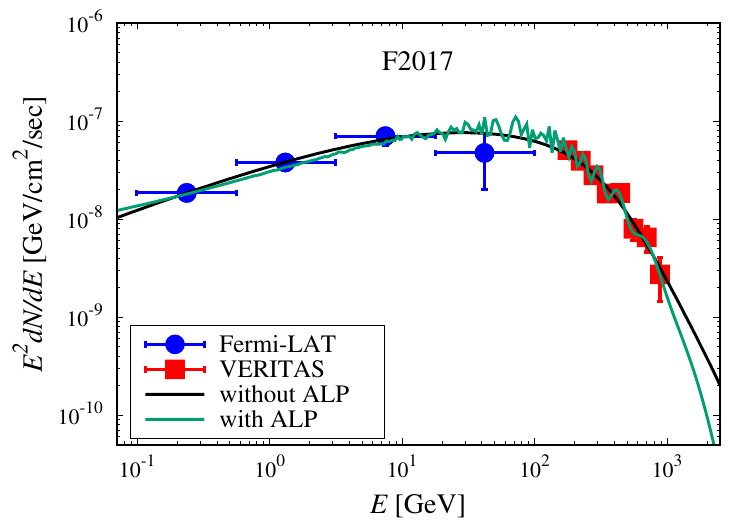} 
\caption{The TeV gamma-ray SEDs of 1ES\,1215+303 with/without the photon-ALP oscillations, corresponding to the flux phases L2008-2012, L2015, L2016, L2017, and F2017.
The black and green lines represent the best-fit SEDs under the null and ALP hypotheses, respectively.
The blue and red experimental data are taken from Fermi-LAT and VERITAS \cite{Fermi-LAT:2020pst}, respectively.
}
\label{fig_SEDs}
\end{figure*}

\section{Long-term gamma-ray observations of 1ES\,1215+303}%%%%%%%%%%%sec_source
\label{sec_source}

In this section, we describe the long-term TeV gamma-ray observations of 1ES\,1215+303.
In Ref.~\cite{Fermi-LAT:2020pst}, the 10-year (2008-08-04 $-$ 2017-09-05) non-flare gamma-ray spectra of 1ES\,1215+303 measured by Fermi-LAT and VERITAS are defined as the flux phases L2008-2012, L2015, L2016, and L2017, respectively.
Note that these four flux phases are low states of the source, and there are still several flares in the observations. 
Since the flares are almost either in GeV or TeV and have a short duration, we just consider one GeV-TeV flare in 2017, which is defined as the flux phase F2017.
See Fig.~\ref{fig_SEDs} for the experimental data of Fermi-LAT and VERITAS.

In this work, the VHE gamma-ray intrinsic spectrum $\Phi_{\rm int}(E)$ with the energy $E$ is selected as the power law with a superexponential cut-off (SEPWL) model, which is described by 
\begin{eqnarray}
\Phi_{\rm int}(E)=N_0\left(E/E_0\right)^{-\Gamma}\exp\left(-\left(E/E_c\right)^d\right)\, ,
\end{eqnarray}
where $N_0$ is a normalized constant, $\Gamma$ is the spectral index, $E_0=1\, \rm GeV$, $E_c$ and $d$ are free parameters. 
Then the chi-square value is given by 
\begin{eqnarray}
\chi_{\rm null}^2 = \sum_{i=1}^{N}\left(\left(\Phi_i - \psi_i\right)/\delta_i\right)^2\, ,
\end{eqnarray}
where $N$ is the gamma-ray spectral point number, $\Phi_i$ is the expected spectrum, $\psi_i$ and $\delta_i$ are the detected flux and its uncertainty, respectively, which is taken from the experimental data.
Since the main effect on VHE photon in the extragalactic space is the EBL photon absorption effect with a factor $e^{-\tau}$, where $\tau$ is the optical depth, then the expected spectrum can be described by
\begin{eqnarray}
\Phi_i=e^{-\tau}\Phi_{\rm int}(E_i)\, .
\end{eqnarray}
In this work, the spectrum of the EBL is taken as the model Franceschini-08 \cite{Franceschini:2008tp}.

Then we show the best-fit TeV gamma-ray SEDs of 1ES\,1215+303 in Fig.~\ref{fig_SEDs} with above five flux phases L2008-2012, L2015, L2016, L2017, and F2017.
The black lines represent the best-fit SEDs under the null hypothesis.
We also list the corresponding best-fit chi-square $\chi^2_{\rm null}$ values in Table~\ref{tab_chi-square}.
Due to the large uncertainty in the observed spectrum of VERITAS, the $\chi^2_{\rm null}$ value of the flux phase L2008-2012 is significantly lower than the other phases.

\begin{table}[t]%%%%%%%%%%%%%table_chi-square
\caption{The best-fit chi-square values under the null and ALP hypotheses of the five flux phases of 1ES\,1215+303.
The combined result is also shown. 
}
\begin{ruledtabular}
\begin{tabular}{lccr}
Flux phase (points) & $\chi^2_{\rm null}$ & $\chi^2_{\rm null}/{\rm d.o.f.}$ & $\chi^2_{\rm min}$ \\
\hline
L2008-2012 (14)   &  3.71    & 0.37  &  2.50 \\
L2015 (14)           &  12.25  & 1.23  &  4.61 \\
L2016 (14)           &  10.44  & 1.04  &  4.29 \\
L2017 (16)           &  6.91    & 0.58  &  2.61 \\
F2017 (12)           &  5.75    & 0.72  &  2.91  \\
\hline
combined             & 39.06   & $\cdots$ & 28.50    \\
\end{tabular}
\end{ruledtabular}
\label{tab_chi-square}
\end{table}

\section{Photon-ALP oscillations}%%%%%%%%%%%%%Photon-ALP oscillations
\label{sec_setup}

\begin{figure}[t]%%%%%%%%%%%%%%%%%%pgg
\centering
\includegraphics[width=0.49\textwidth]{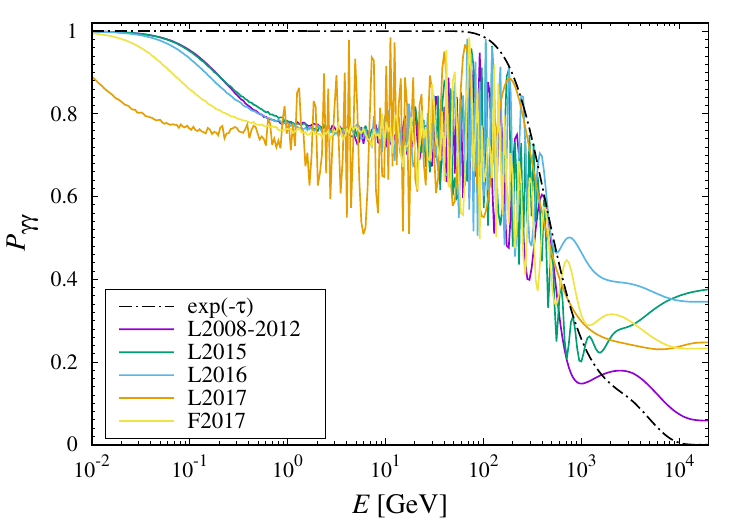}
\caption{The distributions of the final photon-ALP-photon oscillation probability corresponding to $\chi^2_{\rm min}$.
The chain line represents the pure EBL absorption effect with $e^{-\tau}$.
The solid lines represent the final oscillation probability $P_{\gamma\gamma}$ corresponding to $\chi^2_{\rm min}$ in the flux phases L2008-2012, L2015, L2016, L2017, and F2017, respectively.
}
\label{fig_pgg}
\end{figure}

In the following, we consider the photon-ALP oscillation effects on the VHE gamma-ray propagations.
Generally, in the homogeneous magnetic field, the photon-ALP oscillation probability is given by $P_{a\gamma}=(g_{a\gamma}B_T/\Delta_{\rm osc})^2 \sin^2(\Delta_{\rm osc} x_3/2)$, where $B_T$ is the transverse magnetic field, $\Delta_{\rm osc}$ is the oscillation wave number, and $x_3$ is the photon/ALP propagation direction.
See also Refs.~\cite{DeAngelis:2011id, Li:2022mcf}.
When considering the propagation of photon-ALP from the gamma-ray source to the Earth, this process can be divided into three parts, ($a$) the source region, ($b$) the extragalactic space, and ($c$) the Milky Way.
For the propagation distance $s$, the final photon-ALP-photon oscillation probability $P_{\gamma\gamma}$ is 
\begin{eqnarray}
P_{\gamma\gamma}={\rm Tr}\left(\left(\rho_{11}+\rho_{22}\right)\mathcal{T}(s)\rho(0)\mathcal{T}^\dagger(s)\right)\, ,
\end{eqnarray}
where $\mathcal{T}(s)=\mathcal{T}_c(s_c)\times\mathcal{T}_b(s_b)\times\mathcal{T}_a(s_a)$ is the whole transfer matrix, $\rho_{ii}={\rm diag}(\delta_{i1},\delta_{i2},0)$, $\rho(0)={\rm diag}(1,1,0)/2$ is the photon-ALP initial density matrix, and $\rho(s)=\mathcal{T}(s)\rho(0)\mathcal{T}^\dagger(s)$ is the final density matrix. 
 
We first discuss the photon-ALP oscillations in ($a$) the source region of 1ES\,1215+303. 
As the BL Lac object, the blazar jet magnetic field can be described by the poloidal and toroidal components.
We consider a transverse magnetic field model $B(r) = B_0(r/r_{\rm VHE})^{-1}$ and a electron density model $n_{\rm el}(r) = n_0(r/r_{\rm VHE})^{-2}$, where $r_{\rm VHE}$ represents the distance between the source central black hole and the VHE emission region, $B_0$ and $n_0$ correspond to the core magnetic field and electron density at $r_{\rm VHE}$, respectively.
The parameter $r_{\rm VHE}$ is given by $r_{\rm VHE}\simeq R_{\rm VHE}/\theta_{\rm jet}$, with the radius of the VHE emission region $R_{\rm VHE}$, and the angle between the jet axis and the line of sight $\theta_{\rm jet}$.
For the jet region $r > 1\rm\, kpc$, the magnetic field is taken as zero.
The energy transformation between the laboratory and co-moving frames, $E_L$ and $E_j$, should also be considered with the Doppler factor $\delta_{\rm D}=E_L/E_j$.
For the four low state flux phases of the source 1ES\,1215+303 \cite{Fermi-LAT:2020pst}, we take $B_0=2.35\times10^{-2}\, \rm G$, $n_0\simeq1\times10^3\, \rm cm^{-3}$, $R_{\rm VHE}=5.1\times10^{16}\, \rm cm$, $\theta_{\rm jet}=2.0^\circ$, $r_{\rm VHE}=0.47\, \rm pc$, and $\delta_{\rm D}=25$;
While for the one flare flux phase, we take $B_0=5.2\times10^{-2}\, \rm G$, and make other parameters unchanged.
In addition, for the host galaxy region of 1ES\,1215+303, we neglect the photon-ALP oscillations in this region. 

\begin{figure}[t]%%%%%%%%%%%%%%%%%%chi-square 
\centering
\includegraphics[width=0.48\textwidth]{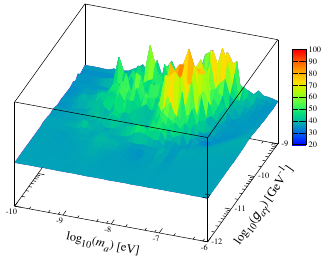}
\caption{The best-fit chi-square distribution in the $\{m_a, \, g_{a\gamma}\}$ plane for the ALP combined analysis.}
\label{fig_chi-square}
\end{figure}

Then in ($b$) the extragalactic space, the main effect on photon-ALP propagation is the EBL absorption effect on VHE photon through the pair-production process.
Due to the weak magnetic field strength in this region, suggesting the upper limit $\sim\mathcal{O}(1)\, \rm nG$ \cite{Ade:2015cva, Pshirkov:2015tua}, here we also neglect the photon-ALP oscillations.

While in ($c$) the Milky Way, we consider the photon-ALP oscillations again in the Galactic magnetic field.
This magnetic field can be modeled as the disk and halo components (both parallel to the Galactic plane), and the ``X-field" component (out-of-plane) at the Galactic center \cite{Jansson:2012pc, Jansson:2012rt, Planck:2016gdp}.

\begin{figure*}[t]%%%%%%%%%%%%%%%%%limit
\centering
\includegraphics[width=0.77\textwidth]{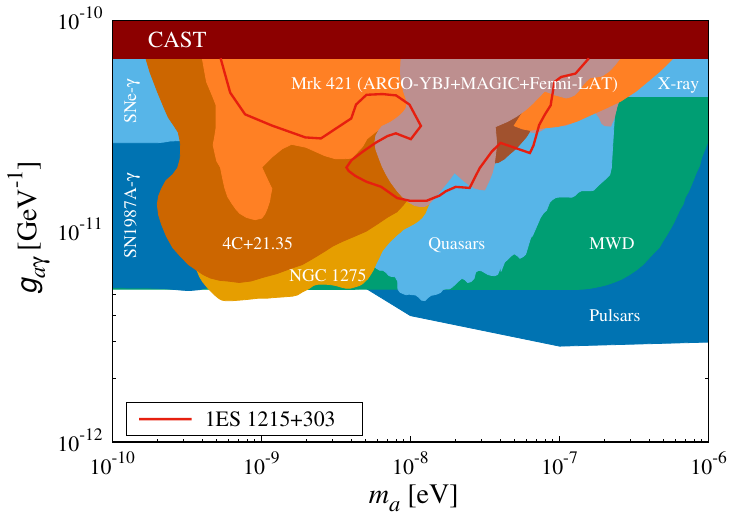}
\caption{The photon-ALP limits in the $\{m_a, \, g_{a\gamma}\}$ plane.
The red contour represents the 95\% $\rm C.L.$ combined limit set by 1ES\,1215+303 with the 10-year gamma-ray observations measured by Fermi-LAT and VERITAS.
See also Ref.~\cite{ciaran_o_hare_2020_3932430} for other limits in this plane. 
}
\label{fig_limit}
\end{figure*}

After considering above photon-ALP oscillation effects, we can derive the final photon-ALP-photon oscillation probability for different ALP parameters $\{m_a, \, g_{a\gamma}\}$.
Then the expected gamma-ray spectrum under the ALP hypothesis is given by 
\begin{eqnarray}
\Phi_{{\rm ALP},\, i} = P_{\gamma\gamma} \Phi_{\rm int}(E_i)\, ,
\end{eqnarray}
where $\Phi_{\rm int}(E_i)$ is the gamma-ray intrinsic spectrum.
In this case $\Phi_i\to\Phi_{{\rm ALP},\, i}$, we have the chi-square value 
\begin{eqnarray}
\chi_{\rm ALP}^2 = \sum_{i=1}^{N}\left(\left(\Phi_{{\rm ALP},\, i} - \psi_i\right)/\delta_i\right)^2\, .
\end{eqnarray}
For one ALP parameters $\{m_a, \, g_{a\gamma}\}$ set, we can derive the corresponding best-fit chi-square $\chi^2_{\rm ALP}$.
Then we can obtain a distribution of the best-fit chi-square in the $\{m_a, \, g_{a\gamma}\}$ plane, in which the minimum chi-square is defined as $\chi^2_{\rm min}$.
We show the best-fit gamma-ray SEDs corresponding to $\chi^2_{\rm min}$ in Fig.~\ref{fig_SEDs} (green lines) with the flux phases L2008-2012, L2015, L2016, L2017, and F2017.
The values of $\chi^2_{\rm min}$ in these phases are listed in Table~\ref{tab_chi-square}.
Compared with the null hypothesis (black lines), the minimum best-fit chi-square under the ALP hypothesis can be significantly depressed.
We also show the distributions of the final photon-ALP-photon oscillation probability corresponding to $\chi^2_{\rm min}$ in Fig.~\ref{fig_pgg}.

Here we set the combined limit on ALP with the five flux phases together.
This is a common method of the ALP analysis for the same source with multiple flux phases \cite{Li:2020pcn}.
We show the best-fit chi-square distribution in the $\{m_a, \, g_{a\gamma}\}$ plane for the ALP combined analysis in Fig.~\ref{fig_chi-square}.
The corresponding best-fit chi-square values are also listed in Table~\ref{tab_chi-square}.

In order to set the 95\% confidence level ($\rm C.L.$) limit on ALP, $\rm i.e.$, to obtain the threshold chi-square $\chi^2_{95\%}$, we simulate 400 sets of the TeV gamma-ray observations of 1ES\,1215+303 in the pseudoexperiments by Gaussian samplings to derive the test statistic (TS) distribution.
The TS obeys the non-central chi-square distribution with the effective $\rm d.o.f.$ and the non-centrality $\lambda$, which is given by ${\rm TS}={\widehat{\chi}_{\rm null}}^2 - {\widehat{\chi}_{\rm ALP}}^2$, where ${\widehat{\chi}_{\rm null}}^2$ and ${\widehat{\chi}_{\rm ALP}}^2$ are the best-fit chi-square of the null and ALP hypotheses in the Monte Carlo simulations, respectively.
Generally, this TS distribution can be approximated with the ALP hypothesis and we use it to derive the 95\% $\rm C.L.$ chi-square difference $\Delta\chi^2_{95\%}$.
Finally, we have the relation $\chi^2_{95\%}=\chi^2_{\rm min}+\Delta\chi^2_{95\%}$, corresponding to the 95\% $\rm C.L.$ threshold chi-square.

In our ALP combined analysis, we have the effective $\rm d.o.f. = 6.71$  with the non-centrality $\lambda=0.01$, suggesting the threshold chi-square $\chi^2_{95\%}=42.15$.
Then we show the 95\% $\rm C.L.$ combined limit result in the $\{m_a, \, g_{a\gamma}\}$ plane, see Fig.~\ref{fig_limit}.
The red contour represents the 95\% $\rm C.L.$ combined limit set by 1ES\,1215+303 with the 10-year gamma-ray observations measured by Fermi-LAT and VERITAS, which is roughly at the photon-ALP coupling constant 
\begin{eqnarray}
g_{a\gamma} \gtrsim 1.5\times 10^{-11} \rm \, GeV^{-1}\, ,
\end{eqnarray}
for the ALP mass 
\begin{eqnarray}
5.0\times10^{-10} \, {\rm eV} \lesssim m_a \lesssim 1.0\times10^{-7} \, {\rm eV}\, .
\end{eqnarray}
The other latest photon-ALP constraints in this plane are also shown, see Ref.~\cite{ciaran_o_hare_2020_3932430} for more details.

Finally, we briefly comment on the impact of parameter uncertainties on the ALP limits.
According to our previous studies \cite{Li:2020pcn, Li:2021gxs}, the parameters that have the greatest impact on the final results are the blazar jet magnetic field parameters $B_0$ and $r_{\rm VHE}$.
Both large $B_0$ and $r_{\rm VHE}$ will lead to more stringent limit.
See also Ref.~\cite{Li:2022jgi}.
In addition, the redshift uncertainty of the source also has an impact, both the underestimated and overestimated redshifts can affect the final results \cite{Li:2022pqa}.

\section{Conclusion}%%%%%%%%%%%%%%%%%%%%%%%Conclusion
\label{sec_conclusion}

In this paper, we have presented the photon-ALP limits from the 10-year period observations of the blazar 1ES\,1215+303.
The long-term TeV gamma-ray spectra are measured by Fermi-LAT and VERITAS with four low states (L2008-2012, L2015, L2016, and L2017) and one flare (F2017) from 2008 to 2017.
For comparison, we show the SEDs of these flux phases under the null and ALP hypotheses.
Then we set the 95\% $\rm C.L.$ combined limit in the ALP parameters $\{m_a, \, g_{a\gamma}\}$ plane.
The 95\% $\rm C.L.$ photon-ALP combined limit set by 1ES\,1215+303 is roughly at the photon-ALP coupling constant $g_{a\gamma} \gtrsim 1.5\times 10^{-11} \rm \, GeV^{-1}$ for the ALP mass $5.0\times10^{-10} \, {\rm eV} \lesssim m_a \lesssim 1.0\times10^{-7} \, {\rm eV}$.
This result is consistent with the limits set by the long-term gamma-ray observations of the source Mrk~421 \cite{Li:2020pcn, Li:2021gxs}.

%\section*{Acknowledgments}%%%%%%%%%%%%%%%%Acknowledgments
{\bf Acknowledgments.} 
We thank Qi Feng, Janeth Valverde, and Olivier Hervet for generously sharing the experimental data of Fermi-LAT and VERITAS with us.
W.C. is supported by the National Natural Science Foundation of China (NSFC) (Grants No.~11775025 and No.~12175027).
Y.F.Z. is supported by the National Key R\&D Program of China (Grant No.~2017YFA0402204), the CAS Project for Young Scientists in Basic Research YSBR-006, and the NSFC (Grants No.~11821505, No.~11825506, and No.~12047503).

\bibliography{references}
\end{document}